\documentclass[11pt,a4paper]{article}
\usepackage{graphicx}
\usepackage{hyperref}
\begin{document}

\title{Generating conjecture and Einstein-Maxwell field of plane symmetry}

\author{J. Fikar \thanks{E--mail: {\tt fikar@physics.muni.cz}}, J. Horsk\'y \thanks{E--mail: {\tt horsky@physics.muni.cz}}\\ Department of Theoretical Physics and Astrophysics\\ Faculty of Science, Masaryk University\\ Kotl\'a\v{r}sk\'a 2, 611 37 Brno, Czech Republic}
\date{19th January 1999}
\maketitle

\begin{abstract}
For the plane symmetry we have found the electro-vacuum exact solutions of the Einstein-Maxwell equations and we have shown that one of them is equivalent to the McVittie solution of a charged infinite thin plane. The analytical extension has been accomplished and the Penrose conformal diagram has been obtained as well.
\end{abstract}

\section{Introduction}
Symmetries and Killing vectors play an important role in exact solutions, especially in the electro vacuum ones e.g. \cite{bicak}. A general plane symmetric metrics in cylindrical coordinates is taken as a seed solution
\begin{equation}
\label{seed}
ds^{2}_{seed}=-e^{2\nu(t,z)}dt^{2}+Y(t,z)^{2}\left(d\rho^{2}+\rho^{2}d\phi^{2}\right)+e^{2\lambda(t,z)}dz^{2},\label{sedm}
\end{equation}
where $\nu(t,z)$, $Y(t,z)$ and $\lambda(t,z)$ are arbitrary functions of coordinates $t$ and $z$.

The generating conjecture, formulated in \cite{horsky1}, \cite{horsky2}, \cite{horsky3}, is applied to azimuthal $\frac{\partial}{\partial\phi}$ Killing vector belonging to rotation in planes of symmetry. Components of this Killing vector are
\begin{equation}
\xi_{a}dx^{a}=\rho^{2}Y(t,z)^{2}d\phi.
\end{equation}

Expected form of ``charged'' metrics is again the general plane symmetric metrics
\begin{equation}
ds^{2}_{new}=-e^{2\alpha(t,z)}dt^{2}+e^{\beta(t,z)}\left(d\rho^{2}+\rho^{2}d\phi^{2}\right)
+e^{2\gamma(t,z)}dz^{2},\label{newm}
\end{equation}
where $\alpha(t,z)$, $\beta(t,z)$ and $\gamma(t,z)$ are functions different
from (\ref{seed}) and they will be evaluated out of the Einstein-Maxwell equations. According to the conjecture we can form an electro-magnetic potential from Killing vector
\begin{equation}
A_\phi=k\rho^{2}Y(t,z)^2.\label{pot}
\end{equation}

\section{Calculations}
If we use the electro-magnetic potential (\ref{pot}) for purely electro-magnetic energy-momentum tensor, two Einstein equations $(t,\rho)$ and $(\rho,z)$ for the space-time (\ref{newm}) are
\begin{eqnarray}
k^{2}\rho^{2}Y(t,z)^{3}\frac{\partial Y(t,z)}{\partial t}e^{-2\beta(t,z)}&=&0, \\
k^{2}\rho^{2}Y(t,z)^{3}\frac{\partial Y(t,z)}{\partial z}e^{-2\beta(t,z)}&=&0.
\end{eqnarray}
These lead to the restrictions on function $Y(t,z)$ in seed metric (\ref{sedm})
\begin{equation}
\frac{\partial Y(t,z)}{\partial t}=0,\ \frac{\partial Y(t,z)}{\partial
z}=0.
\end{equation}

Then the Maxwell equations for (\ref{newm}) are automatically satisfied, the nonzero covariant and contravariant components of the Maxwell electro-magnetic tensor are
\begin{eqnarray}
F_{\rho\phi}&=&-2k\rho Y^{2}, \\
F^{\rho\phi}&=&-\frac{2}{\rho}kY^{2}e^{-4\beta(t,z)}.
\end{eqnarray}

There are other four non-trivial Einstein equations $(t,t)$, $(t,z)$,
$(\rho,\rho)$ and $(z,z)$ with unknown functions
$\alpha(t,z)$, $\beta(t,z)$ and $\gamma(t,z)$ to be solved
\begin{eqnarray}
e^{2(\alpha-\gamma)}\left(2\beta_z\gamma_z
-3\beta_z^2-2\beta_{zz}\right)
+\beta_t^2+2\beta_t\gamma_t&=&4k^2Y^4e^{2\alpha-4\beta},\label{einst1}\\
-\beta_{tz}-\beta_t\beta_z+\beta_t\alpha_z
+\gamma_t\beta_z&=&0,\label{einst2} \\
e^{-2\alpha}\left(\gamma_t\alpha_t-\gamma_{tt}-\beta_t^2-
\gamma_t^2-\beta_{tt}+\alpha_t\beta_t-\gamma_t\beta_t\right)+ \label{einst3}\\
+e^{-2\gamma}\left(\beta_{zz}+\alpha_z\beta_z+\beta_z^2-
\gamma_z\beta_z+\alpha_z^2-\gamma_z\alpha_z+
\alpha_{zz}\right)&=&4k^2Y^4e^{-4\beta},\nonumber\\
-2e^{2(\gamma-\alpha)}\left(\alpha_t\beta_t-2\beta_{tt}-3\beta_t^2\right)-2
\alpha_z\beta_z-\beta_z^2&=&4k^2Y^4e^{2\gamma-4\beta},\label{einst4}
\end{eqnarray}
where subscript denotes partial derivation.

Now we are trying to solve these equations with the assumption that $\beta(t,z)$ is a function of coordinate $z$ only. Consequently equations (\ref{einst2}) and (\ref{einst4}) lead to independence of functions $\alpha(t,z)$ and $\gamma(t,z)$ on coordinate $t$. Therefore all unknown functions $\alpha$, $\beta$ and $\gamma$ now depend on coordinate $z$ only. We should solve two Einstein equations $(t,t)$ (\ref{einst1}) and $(z,z)$ (\ref{einst4})
\begin{eqnarray}
3\beta_z^2+2\beta_{zz}-2\beta_z\gamma_z&=&-4k^2Y^4e^{2\gamma-4\beta},
\label{einst01} \\
\beta_z^2+2\alpha_z\beta_z&=&-4k^2Y^4e^{2\gamma-4\beta}.
\label{einst02}
\end{eqnarray}

From (\ref{einst02}) we evaluate $\gamma(z)$ which is subsequently put into (\ref{einst01}) then we get only one equation for two unknown functions $\alpha(z)$ and $\beta(z)$
\begin{equation}
\alpha_z\beta_{zz}-\alpha_{zz}\beta_z=\beta_z^3+3\alpha_z\beta_z^2+2\alpha_z^2\beta_z.
\label{einst03}
\end{equation}

We can choose one arbitrary function $f(z)$, let us choose it in the form
\begin{equation}
f(z)=\frac{d\beta(z)}{d\alpha(z)}.
\end{equation}

We obtain the functions $\alpha(z)$ and $\beta(z)$
\begin{eqnarray}
\alpha(z)&=&\frac{1}{2}\left(\ln\left(f(z)\right)+\ln\left(f(z)+2\right)-2\ln\left(
f(z)+1\right)+\ln\left(C_1\right)\right),\\
\beta(z)&=&\ln\left(f(z)+1\right)-\ln\left(f(z)+2\right)+\frac{1}{2}\ln\left(C_2\right),
\end{eqnarray}
where $C_1$ and $C_2$ are integration constants. Next, the new metrics components for the solution (\ref{newm}) can be computed
\begin{eqnarray}
g_{tt}&=&-\frac{C_1f(z)\left(f(z)+2\right)}{\left(f(z)+1\right)^2},\label{gsol1}\\
g_{\rho\rho}&=&C_2\left(\frac{f(z)+1}{f(z)+2}\right)^2,\label{gsol2}\\
g_{\phi\phi}&=&C_2\rho^2\left(\frac{f(z)+1}{f(z)+2}\right)^2,\label{gsol3}\\
g_{zz}&=&-\frac{C_2^2\left(f(z)+1\right)^2\left(\frac{df(z)}{dz}\right)^2}
{4k^2Y^4f(z)\left(f(z)+2\right)^5},\label{gsol4}
\end{eqnarray}
where $f(z)$ is an arbitrary function and $C_1$ and $C_2$ are arbitrary constants. The metric tensor must satisfy Hilbert conditions which lead to inequalities
\begin{equation}
C_1<0, \ C_2>0, \ f(z)\left(f(z)+2\right)<0.
\end{equation}

Now we are looking at simplifying resulting metrics tensor using constants substitutions and coordinate transformations
\begin{eqnarray}
C_1&=&-c_1^2,\\
C_2&=&c_2^2,\\
\bar{k}^2&=&\frac{4k^2Y^4}{c_2^4},\\
\bar{\rho}&=&c_2\rho,\\
\bar{t}&=&c_1t,\\
\frac{\bar{z}}{\bar{k}}-1&=&f(z),\label{transf}
\end{eqnarray}
where striped coordinates denote the new ones. The transformation (\ref{transf}) is formally equivalent to the choice of function $f(z)$ as
\begin{equation}
f(z)=\frac{z}{\bar{k}}-1.
\end{equation}
Thus, from the metric tensor we have removed the arbitrary function $f(z)$ which can be interpreted as an arbitrary transformation in coordinate $z$.

The resulting form of the metric tensor is
\begin{equation}
ds^2=-\frac{(k-z)(k+z)}{z^2}dt^2+\left(\frac{z}{k+z}\right)^2
\left(d\rho^2+\rho^2d\phi^2\right)+\frac{z^2}{\left(k-z\right)(k+z)^5}dz^2,\label{solution}
\end{equation}
where we have abandoned the strips and the Hilbert conditions give
\begin{equation}
|z|<k,\ z\neq0.
\end{equation}

Moreover, we can look for solutions beyond Hilbert conditions. They can be gained from Einstein equations (\ref{einst1}), (\ref{einst2}), (\ref{einst3}) and (\ref{einst4}) with the assumption that $\beta(t,z)$ is function only of coordinate $t$. Then equations (\ref{einst1}) and (\ref{einst2}) lead to independence of functions $\alpha(t,z)$ and $\gamma(t,z)$ on coordinate $z$. So all unknown functions $\alpha(t)$, $\beta(t)$ and $\gamma(t)$ depend only on coordinate $t$ now. The resulting metric tensor has the form
\begin{equation}
ds^2=-\frac{(k-t)(k+t)}{t^2}dz^2+\left(\frac{t}{k+t}\right)^2
\left(d\rho^2+\rho^2d\phi^2\right)+\frac{t^2}{\left(k-t\right)(k+t)^5}dt^2.\label{solution2}
\end{equation}
Corresponding Hilbert conditions for this solution are
\begin{equation}
|t|>k.
\end{equation}
The solution (\ref{solution2}) can be obtained directly from the solution (\ref{solution}) exchanging coordinates $t$ and $z$.

\section{Properties of the solutions}
Let us explore some properties of the solutions (\ref{solution}) and (\ref{solution2}) now. Metrics (\ref{solution}) is singular at $z\rightarrow\pm k,0$. Singularities at $z\rightarrow\pm k$ are only coordinate ones, there exist horizons. On the other hand, $z\rightarrow0$ is a real physical singularity, and scalar invariants of the Riemann tensor diverge here as shown
\begin{equation}
I_1 \equiv R^{abcd}R_{abcd} = 8k^4\frac{(k+z)^6}{z^8}\left((k+z)^2+6k^2\right).
\end{equation}

\begin{figure}
\begin{center}
\includegraphics[width=319pt, height=250pt]{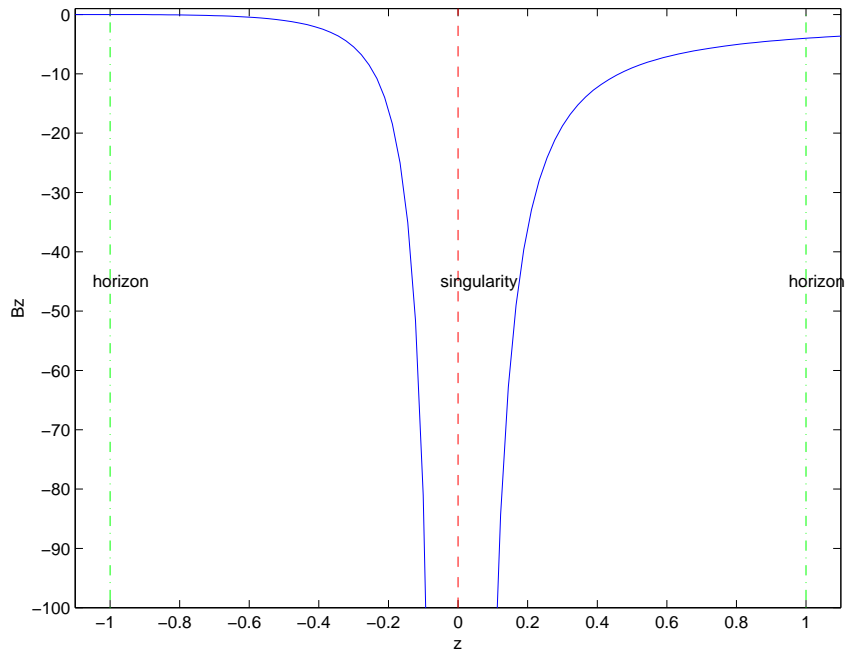}
\end{center}
\caption{Magnetic field}
\label{fig1}
\end{figure}

The electro-magnetic potential in solution (\ref{solution}) is
\begin{equation}
A_\phi=\frac{1}{2}k\rho^2.
\end{equation}
The components of Maxwell electro-magnetic tensor are 
\begin{eqnarray}
F_{\rho\phi}&=&-k\rho, \\
F^{\rho\phi}&=&-\frac{k(k+z)^4}{z^4\rho}.
\end{eqnarray}
We can write down three usual dimensional components of the Maxwell tensor as
\begin{eqnarray}
\vec{E}&=&\left[0,0,0\right], \\
\vec{B}&=&\left[0,0,-k\left(\frac{k+z}{z}\right)^2\right].
\end{eqnarray}
In this solution there exists only a magnetic field. The magnetic field $ \vec{B}$ is shown in Fig.
\ref{fig1} (where $k=1$).

\begin{figure}
\begin{center}
\includegraphics[width=319pt, height=250pt]{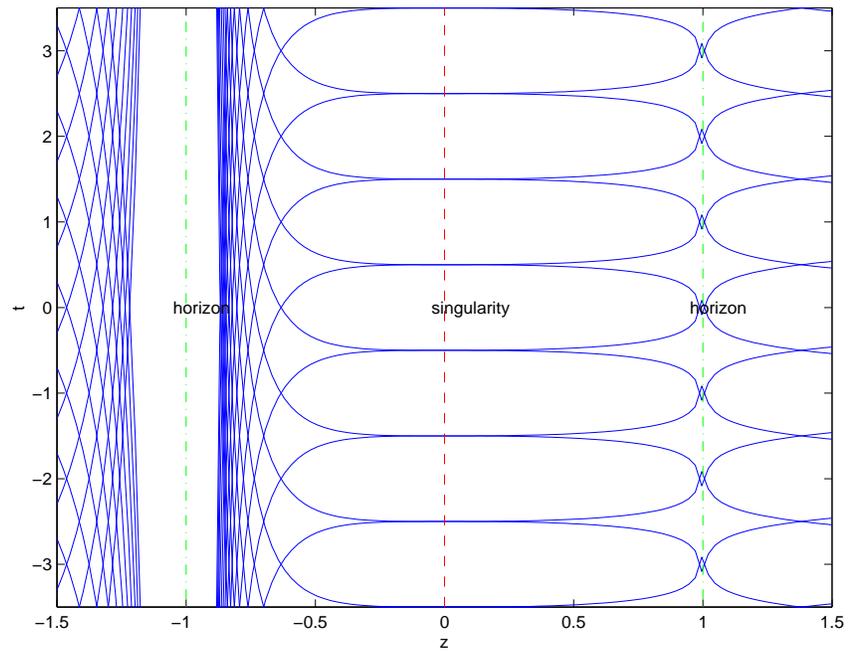}
\end{center}
\caption{Null geodesic lines}
\label{fig2}
\end{figure}

It is also possible to evaluate geodesic trajectories of motion when testing particles are fixed in $t$,$z$ plane. The null geodesic lines are given explicitly by
\begin{equation}
\pm{t(z)}=C+\frac{1}{8k}\ln\left|\frac{k+z}{k-z}\right|+\frac{2k+3z}{4(k+z)^2},
\end{equation}
where $C$ is an integration constant. Such geodesic lines are shown in Fig. \ref{fig2}.

The motion of the charged test particles in the $t$,$z$ plane is given by
\begin{eqnarray}
\ddot{t}+\frac{2k^2\dot{t}\dot{z}}{z\left(z^2-k^2\right)} & = & 0,\\
\ddot{z}+\frac{k^2\left(z+k\right)^4\left(z^2-k^2\right)\dot{t}^2}
{z^5}-\frac{\left(\left(z-k\right)^2+z^2\right)\dot{z}^2}{z\left(z^2-k^2\right)} & = & 0.
\end{eqnarray}
The dots denote differentiation with respect to an affinne parameter. It is clear that the central singular plane acts as a source of repulsive power.

\section{Analytical extension}
Let us consider a static plane-symmetric solution as appears in \cite{gron1} (with null cosmological constant)
\begin{equation}
ds^2=\left(aG^{-1}+bG^{-1/2}\right)dt^2-G\left(dx^2+dy^2\right)-
\frac{G'^2}{4\sigma^2}\left(a+bG^{1/2}\right)^{-1}dz^2.\label{solutiong}
\end{equation}
Where $G$ is an arbitrary function of coordinate $z$, a prime denotes differentiation with respect to $z$, $a$ and $b$ are constants and $\sigma$ is the charge per unit area. When we use the transformations
\begin{eqnarray}
-\frac{4k^2Y^4}{\sigma^2} & = & a+b\sqrt{C_2},\\
0 & = & 2a+b\sqrt{C_2},
\end{eqnarray}
or analogous ones
\begin{eqnarray}
a & = & -C_1C_2,\\
b & = & 2C_1\sqrt{C_2}, \\
\sigma^2 & = & -\frac{C_1C_2}{4k^2Y^4},
\end{eqnarray}
our solution (\ref{gsol1}) - (\ref{gsol4}) turns to be identical with the McVittie (\ref{solutiong}) with $C_2\left(\frac{f(z)+1}{f(z)+2}\right) ^2 \equiv G(z)$. The McVittie solution \cite{mcvitti} is thus generated by the Killing $ \partial /\partial \phi $ vector from the seed metric (\ref{sedm}).

\begin{figure}
\begin{center}
\includegraphics[width=319pt, height=250pt]{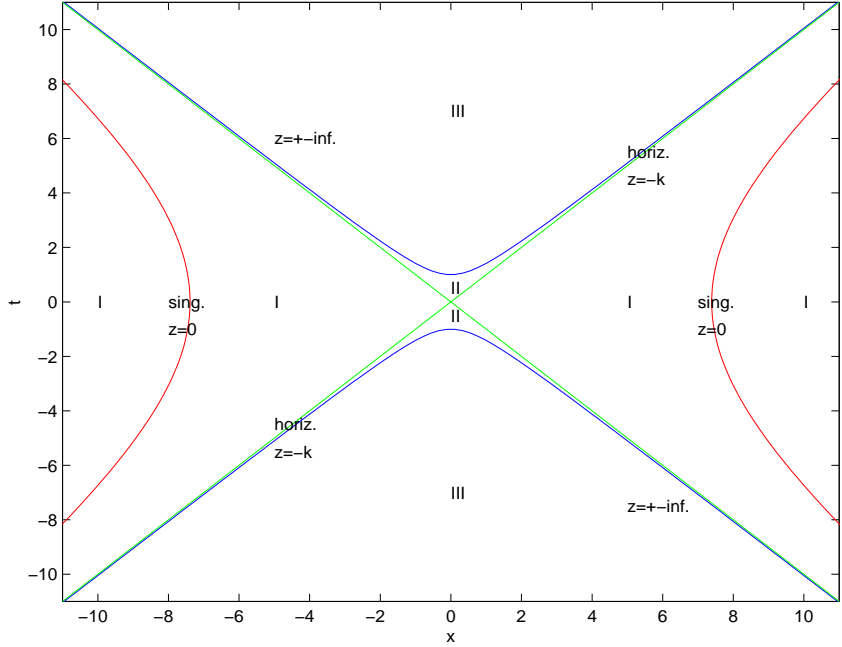}
\end{center}
\caption{Analytical extension}
\label{fig3}
\end{figure}

\begin{figure}
\begin{center}
\includegraphics[width=319pt, height=250pt]{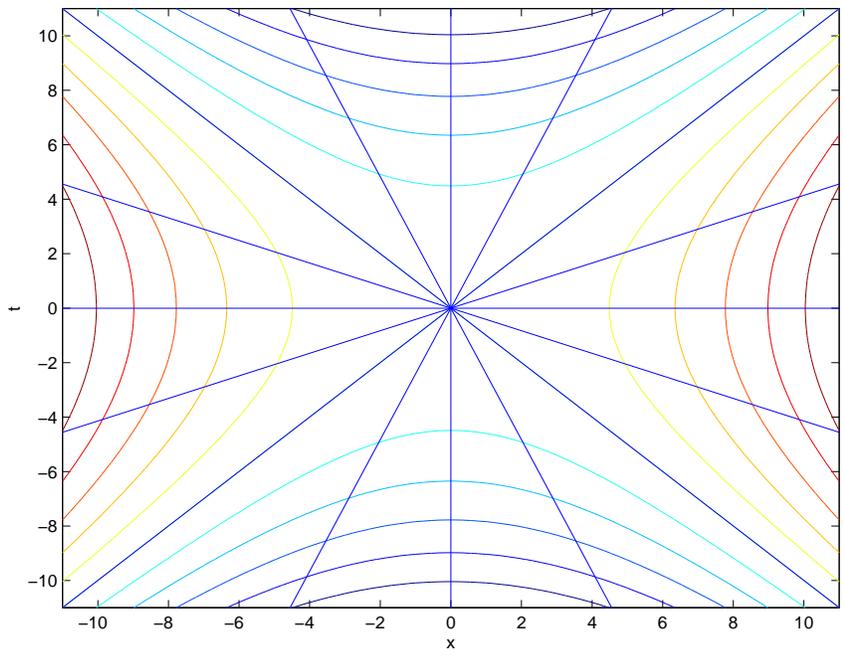}
\end{center}
\caption{New coordinates}
\label{fig4}
\end{figure}

Let us connect both (\ref{solution}) and (\ref{solution2}) solutions together through the horizons $z=\pm k$. We will use retarded and advanced null geodesic coordinates $v$ and $w$ like this
\begin{eqnarray}
v & \equiv & t+\frac{1}{8k}\ln\left|\frac{k+z}{k-z}\right|+\frac{2k+3z}{4(k+z)^2}, \\
w & \equiv & t-\frac{1}{8k}\ln\left|\frac{k+z}{k-z}\right|-\frac{2k+3z}{4(k+z)^2}.
\end{eqnarray}
After transformation of the metric tensor from the cartesian coordinate couple $z$, $t$ to the null one $v$, $w$ we get
\begin{equation}
ds^2 = - \frac{(k-z)(k+z)}{z^2} dvdw+\left(\frac{z}{k+z}\right)^2\left(d\rho^2+\rho^2d\phi^2\right).
\end{equation}
Exponential transformation is used here (which doubles the regions as in Kruskal analytical extension of Schwarzchild solution \cite{hawk} and leaves the coordinates null)
\begin{eqnarray}
v'(v) & = & e^{4kv},\\
w'(w) & = &- e^{-4kw},
\end{eqnarray}
Now again, we diagonalize the metric by using the new cartesian coordinate couple $x'$, $t'$
\begin{eqnarray}
v' & = & t'+x',\\
w' & = & t'-x'.
\end{eqnarray}
This leads to the metrics tensor
\begin{equation}
ds^2=\frac{(k-z)^2e^{2k\frac{2k+3z}{(k+z)^2}}}{16k^2z^2}\left(dx'^2-dt'^2\right)
+\left(\frac{z}{k+z}\right)^2\left(d\rho^2+\rho^2d\phi^2\right),\label{kruskal}
\end{equation}
where $z$, which is $z$ in (\ref{solution}) for $|z|<k$ and $t$ in (\ref{solution2}) for $|z|>k$, is given implicitly by the equation
\begin{equation}
-t'^2+x'^2=\frac{k+z}{k-z}e^{2k\frac{2k+3z}{(k+z)^2}}.
\end{equation}
At first sight there are no sign changes between $z<|k|$ and $z>|k|$, the Hilbert conditions are always satisfied for all values of $x'$, $t'$. The resulting space-time contains every original region I: $|z|<k$, II: $z<-k$, and III: $z>k$ twice as shown in Fig. \ref{fig3}. Furthermore, in (\ref{kruskal}) there are still singularities at $z=0, -k$ but no more at $z=k$. Lines with constant $z$ form over $t'$, $x'$ plane a hyperbolic paraboloid and lines with constant $t$ form lines $t'=const\cdot{x'}$ as shown in Fig. \ref{fig4}. Singular points $z=-k, 0, k,\pm\infty$ are represented by values $0, e^4, \pm\infty,-1$ of $-t'^2+x'^2$, the apparent singularity $z=k$ is moved to infinity in these coordinates.

\begin{figure}
\begin{center}
\includegraphics[width=319pt, height=250pt]{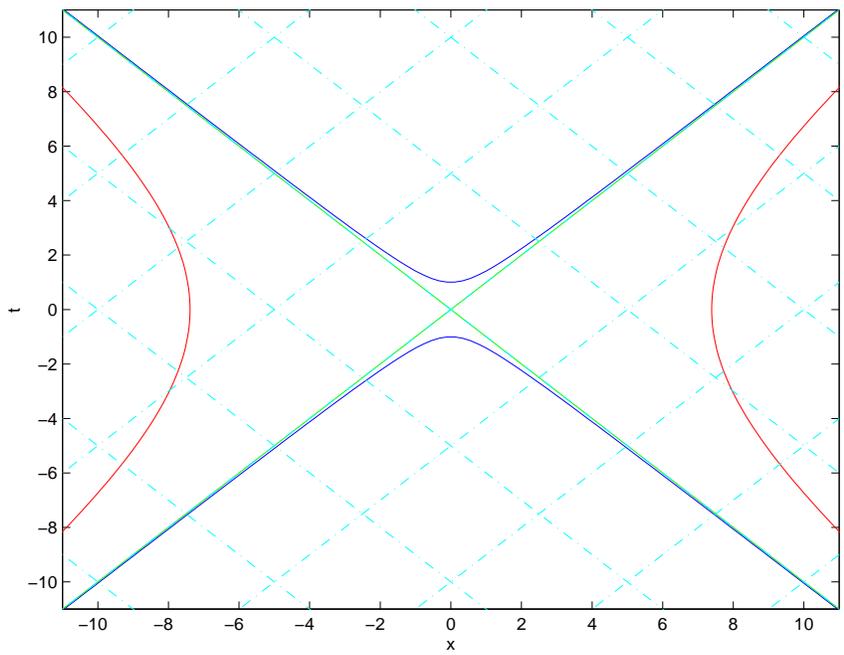}
\end{center}
\caption{Null geodesic lines}
\label{fig5}
\end{figure}

Null geodesic lines in $x',t'$ are straight lines
\begin{equation}
\pm{t'} = x' + const,
\end{equation}
as displayed by dashed lines in Fig. \ref{fig5}.

\begin{figure}
\begin{center}
\includegraphics[width=319pt, height=250pt]{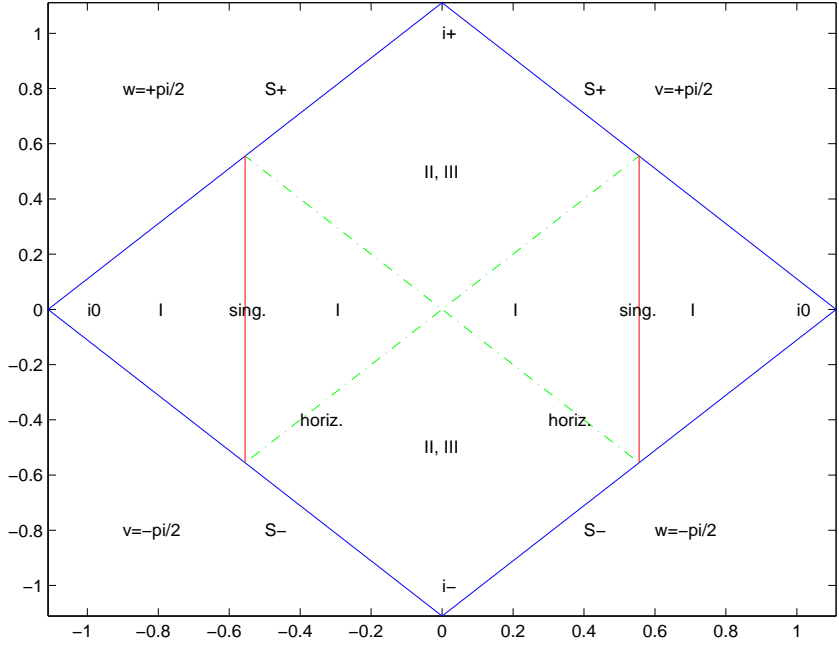}
\end{center}
\caption{Penrose diagram}
\label{fig6}
\end{figure}

The transformation to the Penrose conformal diagram is done by these transformations (again leave null coordinates)
\begin{eqnarray}
v'' & = & \arctan{\left(e^{-2}v'\right)},\\
w'' & = & \arctan{\left(e^{-2}w'\right)}.
\end{eqnarray}
New null coordinates $v'',w''$ are limited by
\begin{eqnarray}
-\frac{\pi}{2}< & v'' & <\frac{\pi}{2},\\
-\frac{\pi}{2}< & w'' & <\frac{\pi}{2},
\end{eqnarray}
so the infinity has been conformaly moved to finite distance. Now it is possible to draw the Penrose diagram (Fig. \ref{fig6}).

Here are marked geodesic future and past $i_+,i_-$, null future and past $S_+, S_-$ and the time-like infinity $i_0$. For the physical singularity $z=0$ is
\begin{equation}
v''-w''=\pm\frac{\pi}{2}
\end{equation}
and both are time-like, so it is possible for an observer to avoid falling in it.

\section{Conclusion}
We have found two electro-vacuum plane symmetric exact solutions of the Einstein-Maxwell equations using the generating conjecture and the Killing vector $ \partial /\partial \phi $ of the plane symmetric vacuum seed metric. We have shown that one of them is equivalent to the McVittie solution of a charged infinite thin plane. We also carried out analytical extension of coordinates, which can cover both found solutions with only one coordinate chart. The corresponding Penrose conformal diagram has been obtained as well.


\begin{thebibliography}{99}
\bibitem{bicak} Bi\v{c}\'ak J., Pravdov\'a A., Symmetries of asymptotically flat electrovacuum spacetimes and radiation, J. Mathem. Phys. 39, 6011 (1998)
\bibitem{horsky1} Horsk\'y J., Mitskievich N. V.: Czech J. Phys. B 39 (1989), page 957
\bibitem{horsky2} Cataldo M., Horsk\'y J., Mitskievich N. V. : Killing vectors and Einstein-Maxwell Field, Differential Geometry and Its Applications, Proc. Conf. Brno (1989)
\bibitem{horsky3} Novotn\'y J., Horsk\'y J. : Generation Method for Solutions of Einstein-Maxwell Equations and Its Applications, Proc. Conf. Opava (1992)
\bibitem{gron1} Amundsen P. A., Gr{\o}n {\O}.: Physical Review D, vol. 27, num. 8 (1983), page 1731
\bibitem{mcvitti} McVittie G. C.:Proc. R. Soc. London 124, 366 (1929)
\bibitem{hawk} Hawking S. W., Ellis G. F. R.: The Large Scale Structure of Space-Time, Cambridge university press (1973)
\end{thebibliography}
\end{document}